\documentclass[12pt]{article}
\usepackage{axodraw}
\usepackage{pstricks}
\usepackage{color}
\usepackage{cite}
\usepackage{array}
\usepackage{epsfig}
\usepackage{amssymb}
\usepackage{graphics,graphpap}
\usepackage{amssymb}
\usepackage{amsmath}
\usepackage{slashed}
\usepackage{dsfont}

\def\eps{\varepsilon}
\def\epe{\varepsilon'/\varepsilon}

\newcommand{\gev}{\, {\rm GeV}}
\newcommand{\mev}{\, {\rm MeV}}

\newcommand{\be}{\begin{equation}}
\newcommand{\ee}{\end{equation}}
\newcommand{\bea}{\begin{eqnarray}}
\newcommand{\eea}{\end{eqnarray}}

\newcommand{\bi}{\begin{itemize}}
\newcommand{\ei}{\end{itemize}}

\newcommand{\vcb}{|V_{cb}|}
\newcommand{\vtd}{|V_{td}|}
\newcommand{\vub}{|V_{ub}|}
\newcommand{\vts}{|V_{ts}|}
\newcommand{\vus}{|V_{us}|}

\def\kpn{K^+\rightarrow\pi^+\nu\bar\nu}

\def\klpn{K_{L}\rightarrow\pi^0\nu\bar\nu}

\usepackage{graphicx}

\medskipamount 

\usepackage{fancyhdr}
\advance \headheight by 3.0truept       

\newlength{\textlength}
\newlength{\overlinelength}

 \def\s#1{\setbox0=\hbox{$#1$}%
   \rlap{\ifdim\wd0>.7em\kern.22\wd0\else\kern.1\wd0\fi /}#1}


\begin{document}

\begin{titlepage}
\begin{flushright}
{FLAVOUR(267104)-ERC-17}
\end{flushright}
\vskip1.2cm
\begin{center}
{\Large \bf \boldmath
On the Correlations between Flavour Observables in Minimal 
$U(2)^3$ Models}
\vskip1.0cm
{\bf
Andrzej J. Buras and Jennifer Girrbach}
\vskip0.3cm
Physik Department, TUM, D-85748 Garching, Germany
\\
TUM-IAS, Lichtenbergstr. 2a, D-85748 Garching, Germany\\
\vskip0.51cm


\vskip0.35cm

{\large\bf Abstract\\[10pt]} \parbox[t]{\textwidth}{
We point out a number of correlations between flavour observables present 
in a special class of models, to be called $MU(2)^3$, with an approximate global
 $U(2)^ 3$ flavour symmetry, constrained by a minimal set of spurions 
governing the breakdown of this symmetry. In this framework only
 Standard Model (SM) operators are relevant in $\Delta F=2$ transitions. 
While the New Physics contributions to $\varepsilon_K$ have the same 
pattern as in
models with constrained Minimal Flavour 
Violation (CMFV), the CP-violation induced by
$B^0_{s,d}-\bar B^0_{s,d}$ mixings can deviate from the one in the SM and CMFV models. 
But these deviations in the $B^0_d$ and $B^0_s$ systems are strictly correlated 
by the $U(2)^3$ symmetry with each other. The most important result of 
our paper is the identification of a stringent triple 
$S_{\psi K_S}-S_{\psi\phi}-\vub$ correlation in this class of models that 
constitutes an important test for them 
and in  this context allows to determine $\vub$ by means 
of precise measurements of $S_{\psi K_S}$ and $S_{\psi\phi}$ with only small 
hadronic uncertainties. We also
find that  $MU(2)^ 3$  
models can in principle accommodate both {\it positive} and
 {\it negative} values of $S_{\psi\phi}$, but in the latter case 
 $\vub$ has to be found in the ballpark of its exclusive determinations 
and the particular   $MU(2)^ 3$ model must provide 
a $20\%$ enhancement of $|\varepsilon_K|$. As in this class of models 
$|\varepsilon_K|$ can only be enhanced, this requirement can  in principle naturally  be 
satisfied, although it depends on the model 
considered. We provide an example of a supersymmetric 
$MU(2)^ 3$ model that appears to satisfy these requirements. 
We summarize 
briefly the pattern of flavour violation 
 in rare $K$ and $B_{s,d}$ decays in $MU(2)^3$ and compare it with the 
one found in CMFV models. 
}

\vfill
\end{center}
\end{titlepage}

\setcounter{footnote}{0}

\newpage

\section{Introduction}
\label{sec:1}

The simplest class of extensions of the Standard Model (SM) are models 
with constrained MFV (CMFV) \cite{Buras:2000dm,Buras:2003jf,Blanke:2006ig}
 that similarly to the SM imply stringent correlations between observables in $K$, $B_d$ and $B_s$ systems, while allowing for significant departures from 
SM expectations. These correlations are consistent with 
present flavour data except possibly for the $\Delta M_{s,d}-\varepsilon_K$ correlation, which appears to experience some tension
\cite{Buras:2012ts}. However,  improved 
lattice calculations of the relevant non-perturbative parameters 
and the reduction of the uncertainty in $\vcb$ that enters $\varepsilon_K$ 
as   $\vcb^4$   are necessary in order to firmly establish this result.
Similar comments apply to the  visible $S_{\psi K_S}-\varepsilon_K$ tension 
\cite{Lunghi:2008aa,Buras:2008nn} within the SM.

Now most models with new sources of flavour and CP-violation can presently 
provide a better simultaneous description  of $\Delta M_{s,d}$, $\varepsilon_K$
and $S_{\psi K_S}$ than achieved within the SM and CMFV models.
However, when no flavour symmetries are present in a given model, 
the pattern of deviations from SM and CMFV expectations is not always 
transparent.

The situation is quite different in possibly simplest non-MFV extensions of 
the SM based on a global $U(2)^3$ flavour symmetry, rather than $U(3)^3$ 
symmetry that governs CMFV \cite{Buras:2000dm} and MFV \cite{D'Ambrosio:2002ex}
models. This class of models studied 
in \cite{Barbieri:2011ci,Barbieri:2011fc,Barbieri:2012uh,Crivellin:2011fb,Crivellin:2011sj,Crivellin:2008mq}\footnote{For earlier discussions of $U(2)$ symmetry see
 \cite{Pomarol:1995xc,Barbieri:1995uv}.}
has several interesting features 
that allow to distinguish it from CMFV and MFV models. As discussed very 
recently in \cite{Barbieri:2012bh} the implications of 
$U(2)^3$ symmetry for flavour physics depends on the way it is broken:
\begin{itemize}
\item
In the so-called minimal $U(2)^3$ models ($MU(2)^3$), analyzed in 
 \cite{Barbieri:2011ci,Barbieri:2011fc,Barbieri:2012uh}
 only the minimal set of spurions is used to break the flavour 
symmetry implying 
 rather stringent relations between various flavour observables. 
\item
When additional spurions are included in the so-called generic 
$U(2)^3$ models ($GU(2)^3$), as done in \cite{Barbieri:2012bh}, 
the implications for flavour physics are less constrained but this is
maybe necessary one day if  $MU(2)^3$ will be falsified by the 
future precise data.
\end{itemize}

Here we would like to reconsider the correlations between flavour 
observables in $MU(2)^3$ models. 
As we will see below with respect to $\Delta F=2$ observables 
all models of this class can be parametrized by 
addition to the SM parameters of only {\it three} new parameters: 
one complex phase 
and two {\it real} and {\it positive definite} parameters of which 
one is restricted to be equal to or larger than unity.

We should already state at the beginning that our goal is not to demonstrate 
that the $MU(2)^3$ models are consistent with the present data because this has 
been already shown in \cite{Barbieri:2011ci,Barbieri:2011fc,Barbieri:2012uh}. 
Partly this is due to existing hadronic, parametric and experimental uncertainties in flavour observables. Yet, these uncertainties will be significantly 
reduced in the coming years and it is of interest to ask how 
 $MU(2)^3$ models would face precision flavour data and the reduction 
of hadronic and CKM uncertainties. In this respect correlations between 
various observables are very important and we would like to exhibit these 
correlations by assuming reduced uncertainties in question. We will 
be more explicit about it in Section~\ref{sec:4}.

Now, the most interesting features 
that allow to distinguish $MU(2)^3$ models from CMFV models are as follows:
\begin{itemize}
\item
New CP-violating phases in $B_d^0-\bar B_d^0$ and 
$B_s^0-\bar B_s^0$ systems are present, thereby allowing for departures of the 
corresponding mixing induced CP-asymmetries $S_{\psi K_S}$ and $S_{\psi\phi}$ 
from the SM and CMFV values. However, very importantly, this appears in 
a correlated manner: the new phases $\varphi_{B_d}$ and $\varphi_{B_s}$ 
are forced by the $U(2)^3$ symmetry to be equal\footnote{To our knowledge
 first discussions 
of this relation can be found  
in \cite{Ligeti:2010ia,Blum:2010mj}.}
\be\label{MAIN1}
\varphi_{B_d}=\varphi_{B_s}.
\ee
\item
The stringent CMFV correlations between K-physics observables and B-physics 
observables are generally absent in these models\footnote{However, in specific models 
of this type some correlations between $K$ and $B$ physics could take 
place.}, allowing among other things to avoid 
the $\Delta M_{s,d}-\varepsilon_K$ tension mentioned above. This also means 
that even if NP effects in $B_s$ and $B_d$ meson system would be found to be 
very small one day, this would not necessarily imply, in contrast to CMFV 
models, small effects in rare $K$ decays like $K^+\to\pi^+\nu\bar\nu$ and 
$K_L\to\pi^0\nu\bar\nu$ and the ratio $\epe$.
\item
On the other hand several relations involving CP-conserving observables in 
$B_d$ and $B_s$ meson systems known from CMFV models and also 
within the $K$ system remain valid, even if 
individual predictions for these observables can significantly differ from 
those found in CMFV scenario. This applies in particular to $\Delta M_{s,d}$ 
and $B_{s,d}\to\mu^+\mu^-$.
\end{itemize}

 Several of these properties have been already identified in
 the extensive work of Barbieri and collaborators \cite{Barbieri:2011ci,Barbieri:2011fc,Barbieri:2012uh}.
Here we would like to point out another correlation  in $MU(2)^3$ flavour 
models, 
 which to our knowledge has not been discussed in the literature 
so far:
\begin{itemize}
\item
The triple correlation:
\be\label{Triple}
S_{\psi K_S}-S_{\psi\phi}-\vub,
\ee
which as we will see below will provide a crucial test of the $MU(2)^3$ 
scenario once the three observables will be precisely known. That is 
once the two of these observables are known, the third one is 
predicted in these models subject to small dependence on the 
angle $\gamma$ and other uncertainties mentioned below.
\item
However, as far as $\Delta F=2$ are concerned, the models which pass this 
test should also simultaneously describe the data for $\varepsilon_K$ 
and $\Delta M_{s,d}$.
\end{itemize}

Presently \cite{Nakamura:2010zzi,Clarke:1429149}                 
\be\label{CPdata}
S_{\psi K_S}^{\rm exp}=0.679\pm0.020, \qquad S_{\psi\phi}^{\rm exp}=0.002\pm0.087
\ee
and improvements from the LHCb, ATLAS and CMS are expected in due time. These results confirm the 
general SM and CMFV expectations that mixing induced CP violation 
in $B_d$ decays is much larger than in $B_s$ decays. 

On the other hand the situation with $\vub$ is unclear at present as 
its  values extracted from exclusive and inclusive semi-leptonic B-decays differ 
significantly from each other: \cite{Laiho:2009eu,Nakamura:2010zzi}
\be\label{exclincl}
\vub_{\rm excl}=(3.12\pm0.26)\times 10^{-3}, \quad 
\vub_{\rm incl}=(4.27\pm0.38)\times 10^{-3}.
\ee
For a recent discussion see \cite{Ricciardi:2012pf}.

Therefore the most sophisticated 
numerical analyses of NP effects in various extensions of the 
SM found in the literature use as input some kind of an average between the inclusive and exclusive determination 
of $\vub$. For instance the authors of \cite{Barbieri:2011ci}  use
\be\label{average}
\vub=(3.97\pm0.45)\times 10^{-3}.
\ee
While such an approach is legitimate, it may hide some interesting 
physics behind the values of $\vub$ which are washed out in such an approach. 
Similar comments apply to the analyses in \cite{Altmannshofer:2011gn,Altmannshofer:2012az,Botella:2012ju}. 

In this context we should emphasize that one
of the recent highlights in particle physics was the measurement of 
the angle $\theta_{13}$ in the PMNS matrix with a precision significantly 
larger than the corresponding precision on $\vub$. The unexpectedly high value of $\theta_{13}$ had a large
impact on the models for the PMNS matrix and the field of lepton flavour 
violation. For recent reviews see \cite{Altarelli:2012ss,Altarelli:2012bn}.

Now, as assured by various documents 
on future determinations of $\vub$ 
\cite{Bona:2007qt,Antonelli:2009ws,Ciuchini:2011ca}, the second half of this decade could 
bring also significant progress on $\vub$ and one can anticipate that this 
will also have an impact on the physics of quarks.
Therefore, we have emphasized in recent papers \cite{Blanke:2011ry,Buras:2011wi,Buras:2012ts} that the future precise determination 
of $\vub$ combined with a more accurate determination of the angle $\gamma$ 
and reduced uncertainties in the hadronic parameters relevant for $\Delta F=2$ 
transitions by future lattice calculations, could already on the basis of a few $\Delta F=2$ observables distinguish uniquely between the simplest extensions 
of the SM. In this context we have considered two scenarios for $\vub$: 
 \begin{itemize}
\item
{\bf Exclusive (small) $\vub$ Scenario 1:}
$|\varepsilon_K|$ is smaller than its experimental determination,
while $S_{\psi K_S}$ is rather close to the central experimental value.
\item
{\bf Inclusive (large) $\vub$ Scenario 2:}
$|\varepsilon_K|$ is consistent with its experimental determination,
while $S_{\psi K_S}$ is significantly higher than its  experimental value.
\end{itemize}

Thus dependently which scenario is considered we need either  
{\it constructive} NP contributions to $|\varepsilon_K|$  
 (Scenario 1) 
or {\it destructive} NP contributions to  $S_{\psi K_S}$ (Scenario 2). 
However this  NP should not spoil the agreement with the data 
for $S_{\psi K_S}$ (Scenario 1) and $|\varepsilon_K|$ (Scenario 2).

While introducing these two scenarios, one should emphasize the following difference between them. 
In Scenario 1, the central value of $|\varepsilon_K|$ is visibly smaller than 
the very precise data  but the still  significant parametric uncertainty 
due to $\vcb^4$ dependence in $|\varepsilon_K|$ and a large uncertainty 
in the charm contribution found at the NNLO level in \cite{Brod:2011ty} 
does not make this problem as pronounced as this is the case of 
Scenario 2, where large $\vub$ implies definitely a value of $S_{\psi K_S}$ 
that is by $3\sigma$ above the data.

In the present paper we will generalize these considerations by including 
the full range  
\be\label{Vubrange}
2.8\times 10^{-3}\le\vub\le 4.6\times 10^{-3}
\ee
in our analysis and calculating 
how the correlation between $S_{\psi K_S}$ and $S_{\psi\phi}$ in 
$MU(2)^3$ models depends on the value of $\vub$, which hopefully will 
be known precisely one day. 

Thus the main new result in the present paper is Fig.~\ref{fig:SvsS}, 
which will allow us in the future to monitor how $MU(2)^3$ models 
are facing the improved data on $S_{\psi K_S}$, $S_{\psi\phi}$ and 
$\vub$. We would like to emphasize that hadronic uncertainties 
in  $S_{\psi K_S}$ and $S_{\psi\phi}$ are presently significantly smaller than 
in $\Delta M_{s,d}$ and $\varepsilon_K$, although we are  aware of the possibility that in the era of precision flavour physics additional effects from 
QCD-penguin effects should be taken into account in the extraction of model parameters from 
these asymmetries
\cite{Fleischer:1999nz,Ciuchini:2005mg,Ciuchini:2011kd,Faller:2008zc,Faller:2008gt,Jung:2009pb,DeBruyn:2010hh,Lenz:2011zz,
Boos:2004xp,Li:2006vq, Gronau:2008cc,Jung}. 

Another new result in the context of $MU(2)^3$ models, that follows from 
 Fig.~\ref{fig:SvsS}, is the realization that in these models $S_{\psi\phi}$ 
can not only be smaller than the SM value but also have opposite sign. 
Here we differ from the authors of \cite{Barbieri:2011ci,Barbieri:2011fc} 
who using the $\vub$ in (\ref{average}) concluded that a striking prediction 
of the $U(2)^3$ framework is a value of $S_{\psi\phi}$ in the range 
$0.05\le S_{\psi\phi}\le 0.20$ implying that a future {\it negative} value 
of $S_{\psi\phi}$ would rule out this framework\footnote{The same 
conclusion is reached in the ${\rm 2HDM_{\overline{MFV}}}$ framework 
\cite{Buras:2010mh,Buras:2010zm,Buras:2012ts,Buras:2012xxx} but in 
this case no escape through a low value of $\vub$ is possible as in 
this model
the contributions to $\varepsilon_K$ are tiny.}. Fortunately as  Fig.~\ref{fig:SvsS} demonstrates this is not the case.
However this requires a rather low value of $\vub$, in the ballpark 
of exclusive determinations, and consequently a positive contribution 
to $|\varepsilon_K|$ of the order of $20\%$, which likely cannot be 
achieved in all $MU(2)^3$ models. On the other hand an enhancement of 
 $|\varepsilon_K|$ similarly to CMFV is a unique, basically model 
independent, prediction of the
 $MU(2)^3$ framework.

Another related correlation in this framework 
is the one between $\mathcal{B}(B^+\to\tau^+\nu_\tau)$ and  $S_{\psi\phi}$ 
and given in Fig.~\ref{fig:Btaunu}  
that in due time will constitute an important constraint 
on $MU(2)^3$ models. 
We also point out that in an $MU(2)^3$ flavour symmetric world one 
can determine $\vub$ by means 
of precise measurements of $S_{\psi K_S}$ and $S_{\psi\phi}$ with only small 
hadronic uncertainties. 

 Finally we  point out that 
$B_{s,d}\to\mu^+\mu^-$ decays allow to 
test new CP-violating phases in $MU(2)^3$ models with the help of mixing 
induced 
CP-asymmetries $S^{s,d}_{\mu^+\mu^-}$ proposed in 
\cite{deBruyn:2012wj,deBruyn:2012wk}.

 Our paper is organized as follows. In Section~\ref{sec:2} we describe 
 the departures from SM predictions for $\Delta F=2$ processes in terms 
 of general expressions. In Section~\ref{sec:3} we specify these expressions 
 to CMFV models and $MU(2)^3$ models.
In Section~\ref{sec:4} we present a general numerical 
 analysis of $\Delta F=2$ observables in $MU(2)^3$ models with the main 
results represented by  Figs.~\ref{fig:SvsS}--\ref{fig:Btaunu} 
 and the implications thereof.
In Section~\ref{sec:5} we outline a strategy for testing the 
$MU(2)^3$ scenario in the coming years. Subsequently we execute this 
strategy in the case of a specific
supersymmetric model of the
 $MU(2)^3$ type. While a detailed sophisticated analysis 
 of this model with the $\vub$ in (\ref{average}) has been already 
 presented in \cite{Barbieri:2011ci}, we show that a more specific look at the values 
 of $\vub$ allows to obtain some insight which goes beyond the 
 results obtained by these authors.  In Section~\ref{sec:6} we summarize 
briefly the pattern of flavour violation 
 in rare  $K$ and $B_{s,d}$ decays in $MU(2)^3$  models. 
 A summary of our main results and a brief outlook for the future
 are given in  Section~\ref{sec:7}.

\boldmath
 \section{Basic Formulae for $\Delta F=2$ Observables}
\unboldmath
 \label{sec:2}

The expressions for 
the off-diagonal elements $M^i_{12}$ in the neutral $K$ and $B_{q}$ meson mass matrices for  the models considered in our paper can be
written 
in a general form  as follows
\bea\label{eq:3.4}
\left(M_{12}^K\right)^*&=&\frac{G_F^2}{12\pi^2}F_K^2\hat
B_K m_K M_{W}^2\left[
\lambda_c^{2}\eta_1x_c +\lambda_t^{2}\eta_2S_K +
2\lambda_c\lambda_t\eta_3S_0(x_c,x_t)
\right]
\,,\\
\left(M_{12}^q\right)^*&=&{\frac{G_F^2}{12\pi^2}F_{B_d}^2\hat
B_{B_d}m_{B_d}M_{W}^2
\left[
\left(\lambda_t^{(q)}\right)^2\eta_B S_q
\right]}\,,\label{eq:3.6}
\eea
where $q=d,s$, $x_i=m_i^2/M_W^2$ and
\be
\lambda^{(K)}_i=V_{is}^*V_{id},\qquad \lambda_t^{(q)}=V_{tb}^*V_{tq}
\ee
 with $V_{ij}$ being the elements of the CKM matrix. Here, $S_0(x_c,x_t)$
is a {\it real valued} one-loop box function for which explicit expression is given e.\,g.~in \cite{Blanke:2006sb}. The
factors $\eta_i$ are QCD {corrections} evaluated at the NLO level in
\cite{Herrlich:1993yv,Herrlich:1995hh,Herrlich:1996vf,Buras:1990fn,Urban:1997gw}. For $\eta_1$ and $\eta_3$ also NNLO corrections
have been recently
calculated \cite{Brod:2010mj,Brod:2011ty}. Finally $\hat B_K$ and $\hat B_{B_q}$ are the well-known
non-perturbative factors.

 In the $MU(2)^3$ framework of \cite{Barbieri:2011ci}, similarly to MFV, NP enters these expressions only in terms proportional to
$\lambda^{(K)}_t$ and $\lambda^{(q)}_t$, that is through the functions  $S_i$ with $i=K,q$. But this 
time these functions can be complex:
\be
\label{eq31}
S_i\equiv|S_i|e^{i\theta_S^i}.
\ee
Note that the flavour dependence enters the mixing 
amplitudes through the elements of the CKM matrix and the functions $S_i$ in 
question. We emphasize the complex conjugation in the expressions for 
mixing amplitudes, which is essential as $S_i$ beyond the SM and CMFV carry in principle new 
complex phases.

The $\Delta B=2$ mass differences can now be written as follows:

\be\label{DMd}
\Delta M_d=2|M_{12}^d| =\frac{G_F^2}{6 \pi^2}M_W^2 m_{B_d}|\lambda_t^{(d)}|^2   F_{B_d}^2\hat B_{B_d} \eta_B |S_d|\,,
\ee
\be\label{DMs}
\Delta M_s =2|M_{12}^s|=\frac{G_F^2}{6 \pi^2}M_W^2 m_{B_s}|\lambda_t^{(s)}|^2   F_{B_s}^2\hat B_{B_s} \eta_B |S_s|\,.
\ee
The corresponding mixing induced CP-asymmetries are then given
by
\begin{equation}
S_{\psi K_S} = \sin(2\beta+2\varphi_{B_d})\,, \qquad
S_{\psi\phi} =  \sin(2|\beta_s|-2\varphi_{B_s})\,,
\label{eq:3.44}
\end{equation}
where the phases $\beta$ and $\beta_s$ are defined by 
\be\label{vtdvts}
V_{td}=\vtd e^{-i\beta}, \qquad V_{ts}=-\vts e^{-i\beta_s}.
\ee
$\beta_s\simeq -1^\circ\,$.
 The new phases $\varphi_{B_q}$  are directly related to the phases of the functions
$S_q$:
\be
2\varphi_{B_q}=-\theta_S^q.
\ee

For the CP-violating parameter $\varepsilon_K$ we have
\be
\varepsilon_K=\frac{\kappa_\eps e^{i\varphi_\eps}}{\sqrt{2}(\Delta M_K)_\text{exp}}\left[\Im\left(M_{12}^K\right)\right]\,,
\label{eq:3.35}
\ee
where $\varphi_\eps = (43.51\pm0.05)^\circ$ and $\kappa_\eps=0.94\pm0.02$ \cite{Buras:2008nn,Buras:2010pza} takes into account 
that $\varphi_\eps\ne \tfrac{\pi}{4}$ and includes long distance  effects in $\Im( \Gamma_{12})$ and $\Im (M_{12})$.

In the rest of the paper, unless otherwise stated, we will assume that all four parameters in the CKM 
matrix have been determined through tree-level decays without any NP pollution 
and pollution from QCD-penguin diagrams so that their values can be used 
universally  in  
all NP models considered by us. 

\boldmath
\section{CMFV vs. $MU(2)^3$ Models: $\Delta F=2$ Observables}
\unboldmath
\label{sec:3}
The $\Delta F=2$ observables discussed above can be calculated in 
any model by specifying the functions $S_i$\footnote{Note, that this 
statement applies also to models with new operators as their contributions 
can always be included in the definition of $S_i$ but then these functions 
will depend on new non-perturbative parameters.}. Before doing it here, 
let us summarize briefly the basics of these two classes of extensions 
of the SM.
\subsection{Constrained Minimal Flavour Violation (CMFV)} 
This is possibly the simplest class of BSM scenarios. It is defined 
pragmatically as follows \cite{Buras:2000dm}:
\begin{itemize}
\item
The only source of flavour and CP violation is the CKM matrix. This 
implies that the only CP-violating phase is the KM phase and that
CP-violating flavour blind phases are assumed to be absent.
\item
The only relevant operators in the effective Hamiltonian below 
the electroweak scale are the ones present within the SM.
\end{itemize}
Detailed expositions of phenomenological consequences of this NP 
scenario has been given in \cite{Buras:2003jf,Blanke:2006ig} and recently 
in \cite{Buras:2012ts}. The main implications of this framework  
are listed below.

In the grander formulation by means of effective theory approach 
\cite{D'Ambrosio:2002ex}, CMFV corresponds to the case of one Higgs doublet 
and the pragmatic assumption that no new operators beyond those present in the 
SM at the electroweak scale are relevant. This assumption is useful as 
finding the departures from CMFV could also point out to new operators and/or 
new sources of flavour violation.
\boldmath
\subsection{Minimal $U(2)^3$ Models}
\unboldmath
A pragmatic  definition of these models  in the case of $\Delta F=2$ 
transitions  could be  as follows:

\begin{itemize}
\item
Flavour and CP-violation in the  $\Delta S=2$  transitions is governed by CMFV.
\item
The dominant source of flavour and CP violation in $B_{d,s}-\bar B_{d,s}$ 
mixings  
is the CKM matrix.
Yet, new universal, with respect to $B_d$ and $B_s$, 
flavour violating and CP-violating effects in these  
transitions are possible. The universality in question is a direct 
consequence of the $U(2)^3$ symmetry imposed on the quark doublets of the 
first two generations. Moreover, only
SM operators are relevant in $B_{d,s}-\bar B_{d,s}$ 
mixings.  We comment on the possible small non-universal corrections and 
contributions of new operators below.
\item
Very importantly NP effects in  $K$ physics and $B_{d,s}$ observables 
are  uncorrelated with each other, although in specific models such 
correlation could be forced by the underlying theory and the data.
\end{itemize}

In the grander formulation by means of effective theory \cite{Barbieri:2011ci}
these models are governed by a global flavour symmetry
\be
G_F=U(2)_Q\times U(2)_u\times U(2)_d
\ee
broken {\it minimally} by three spurions transforming under $G_F$ as 
follows
\be
\Delta Y_u=(2,\bar 2,1), \quad \Delta Y_d=(2,1,\bar 2), \quad V=(2,1,1).
\ee

 As demonstrated by means of a spurion analysis in Section 5 of 
\cite{Barbieri:2011ci}, the phenomenological consequences of this 
framework for $\Delta F=2$ transitions as summarized in the pragmatic 
definition are general consequences of $U(2)^3$ symmetry and its breaking 
pattern.  They go beyond  supersymmetry that otherwise dominates 
this paper.
In particular, if one considers leading flavour-changing amplitudes no assumption of the dominance of SM operators has 
to be made. Moreover, the universality of NP effects in $B_{d,s}-\bar B_{d,s}$ 
systems, up to the overall usual CKM factors and the hermicity of the 
$\Delta F=2$ Hamiltonian implies that $S_K$ is real \cite{Barbieri:2011ci}. 
This result in 
combination with the dominance of SM operators in this framework implies 
the CMFV structure of $\Delta S=2$ transitions.

In this context the following remark should be made. As in the MFV framework the leading flavour-changing amplitudes in the framework of  \cite{Barbieri:2011ci} are of left-handed type and to a very good approximation can be evaluated neglecting the effects of light-quark masses. In order to generate dimension six 
LR operators contributing to $\Delta F=2$ transitions one needs at least two extra insertion of the down-type spurion 
$\Delta Y_d$ which implies an extra suppression of amplitudes proportional 
to down-quark masses. Such effects being proportional to light quark masses 
($m_{s,d}$) break the flavour universality between $B_d$ and $B_s$ systems 
and introduce corrections to the relation (\ref{MAIN1}). They can 
be at best relevant for $B_s$-mixing, for instance in 2HDM models or supersymmetric models. However, unless one goes to a specific regime of large $\tan\beta$ 
and small Higgs masses such effects are always subleading and we will neglect them. 

More interesting in this context are the operators 
\be\label{QSLL}
{Q}_1^\text{SLL}=\left(\bar b P_L s\right)\left(\bar b P_L s\right)\,, 
{Q}_2^\text{SLL}=\left(\bar b \sigma_{\mu\nu} P_L s\right)\left(\bar b\sigma^{\mu\nu}  P_L s\right)\,
\ee
present in the ${\rm 2HDM_{\overline{MFV}}}$ framework 
\cite{Buras:2010mh,Buras:2010zm,Buras:2012ts,Buras:2012xxx}. Having 
Wilson coefficients proportional to $m_b^2$, their contributions have the 
same impact on $B_s$ and $B_d$ mixings and satisfy in particular the
relation $(\ref{MAIN1})$. In \cite{Buras:2010zm} they correspond to the 
dominance of flavour blind phases in the Higgs potential, rather than to such
phases in Yukawa couplings corresponding to LR operators and considered 
in \cite{Buras:2010mh}. In fact our statements on  ${\rm 2HDM_{\overline{MFV}}}$ 
in the present paper apply only to the case of the dominance of the 
operators in (\ref{QSLL}) as only in this case the  ${\rm 2HDM_{\overline{MFV}}}$ 
framework has a chance to remove the $\varepsilon_K-S_{\psi\phi}$ tension in the 
SM. However, to generate such operators in the ${\rm 2HDM_{\overline{MFV}}}$  
framework requires not only flavour blind phases but also sizable $SU(2)_L$ 
breaking in the Higgs potential (splitting between $m_A$ and $m_H$).
Finally the contributions of the operators in (\ref{QSLL}) to $\Delta S=2$ 
are proportional to $m_s^2$ and negligible.

\subsection{Comparison}
For our purposes it will be sufficient to discuss first only those general 
properties of the $S_i$ in these two NP scenarios 
from which correlations between various 
observables automatically follow. We have then:

{\bf 1.} In CMFV we have 
\be\label{G1}
S_K=S_d=S_s\ge S_0(x_t), \qquad \varphi_K=\varphi_{B_d}=\varphi_{B_s}=0, \qquad 
({\rm CMFV})
\ee
where $S_0(x_t)$ is the SM box function given by 
\be
S_0(x_t)  = \frac{4x_t - 11 x_t^2 + x_t^3}{4(1-x_t)^2}-\frac{3 x_t^2\log x_t}{2
(1-x_t)^3}~
\ee
and the inequality in (\ref{G1}) has been demonstrated diagrammatically 
in \cite{Blanke:2006yh}.
Consequently 
\be
S_{\psi K_S} = \sin(2\beta)\,,\qquad S_{\psi\phi} =  \sin(2|\beta_s|)\,
\qquad 
({\rm CMFV}).
\label{CMFV1}
\ee
$|\varepsilon_K|$, $\Delta M_d$ and $\Delta M_s$ 
can only be enhanced in CMFV models. Moreover, this happens 
in a correlated manner. The enhancement of one of these observables implies 
automatically and uniquely the enhancement of the other two observables 
\cite{Buras:2000xq,Blanke:2006ig}. 

{\bf 2.} On the other hand in $MU(2)^3$ models (\ref{G1}) and (\ref{CMFV1}) 
are replaced by 
\be\label{G2}
S_K=r_K S_0(x_t), \quad   |S_d|=|S_s|=r_B S_0(x_t), \quad \varphi_K=0,\quad  \varphi_{B_d}=\varphi_{B_s}\equiv\varphi_{\rm new}, \qquad 
({\rm MU(2)^3})
\ee
and consequently 
\begin{equation}
S_{\psi K_S} = \sin(2\beta+2\varphi_{\rm new})\,, \qquad
S_{\psi\phi} =  \sin(2|\beta_s|-2\varphi_{\rm new})\,,\qquad 
({\rm MU(2)^3})
\label{U21}
\end{equation}
implying a correlation between these two asymmetries. As there is no relation 
of $S_d=S_s$ to the SM box function $S_0(x_t)$ and $S_K$ in these models, $\Delta M_d$ and $\Delta M_s$ can be suppressed or enhanced with respect to the SM values and there is no direct correlation between 
 $|\varepsilon_K|$ and  $\Delta M_{s,d}$.

In short, with respect to $\Delta F = 2$ processes there are only three new parameters in this class of models
\be\label{CU3par}
r_K\ge 1, \qquad r_B,\qquad  \varphi_{\rm new}
\ee
with $r_K$ and $r_B$ being real and positive definite.

The first inequality in (\ref{CU3par}) is  a direct consequence 
of the CMFV structure in the $\Delta S=2$ transitions in this framework 
\cite{Blanke:2006yh}. Therefore in the $MU(2)^3$ framework  
 $|\varepsilon_K|$ can only be increased over the SM value, which is 
supported by the data 
but as stated above, this property  is generally 
uncorrelated with $B_{s,d}$ systems.

{\bf 3.}
The flavour universality of the functions $S_q$ in CMFV and $MU(2)^3$ models
implies 
\be\label{CMFV3}
\left(\frac{\Delta M_d}{\Delta M_s}\right)_{\rm CMFV}=
\left(\frac{\Delta M_d}{\Delta M_s}\right)_{\rm MU(2)^3}=
\left(\frac{\Delta M_d}{\Delta M_s}\right)_{\rm SM}=
\frac{m_{B_d}}{m_{B_s}}
\frac{\hat B_{d}}{\hat B_{s}}\frac{F^2_{B_d}}{F^2_{B_s}}
\left|\frac{V_{td}}{V_{ts}}\right|^2
\equiv\frac{m_{B_d}}{m_{B_s}}
\frac{1}{\xi^2}
\left|\frac{V_{td}}{V_{ts}}\right|^2.
\end{equation}

The $\Delta F=1$ observables are discussed in Section~\ref{sec:6}.

\section{General Numerical Analysis}\label{sec:4}
\subsection{Strategy}

As we stated at the beginning of our paper it is not the goal of this section to present a full-fledged numerical 
analysis of all correlations including present theoretical and experimental 
uncertainties as this would only wash out the effects we want to emphasize. 
 We think that in view of the flavour precision era ahead of us it is 
more important to identify certain characteristic features of this NP 
scenario by decreasing present hadronic and parametric uncertainties.
Therefore, in our numerical analysis we will  choose 
as nominal values for three out of four CKM parameters:\vspace{1ex}
\be\label{fixed}
\vus=0.2252, \qquad \vcb=0.0406, \qquad \gamma=68^\circ, 
\ee
where the values for
 $|V_{us}|$ and  $|V_{cb}|$ have been measured
 in tree level
decays. The value for $\gamma$ is consistent with CKM fits and as the 
ratio $\Delta M_d/\Delta M_s$ in the model considered equals the SM one, this 
choice is a legitimate one. Indeed the SM value for this ratio agrees well with 
the data.
Other inputs are collected in
Table~\ref{tab:input}. For $\vub$ we will use the range in (\ref{Vubrange}).

\begin{table}[!tb]
\center{\begin{tabular}{|l|l|}
\hline
$G_F = 1.16637(1)\times 10^{-5}\gev^{-2}$\hfill\cite{Nakamura:2010zzi} 	&  $m_{B_d}= 5279.5(3)\mev$\hfill\cite{Nakamura:2010zzi}\\
$M_W = 80.385(15) \gev$\hfill\cite{Nakamura:2010zzi}  								&	$m_{B_s} =
5366.3(6)\mev$\hfill\cite{Nakamura:2010zzi}\\
$\sin^2\theta_W = 0.23116(13)$\hfill\cite{Nakamura:2010zzi} 				& 	$F_{B_d} =
(190.6\pm4.7)\mev$\hfill\cite{Laiho:2009eu}\\
$\alpha(M_Z) = 1/127.9$\hfill\cite{Nakamura:2010zzi}									& 	$F_{B_s} =
(227.7\pm5.0)\mev$\hfill\cite{Laiho:2009eu}\\
$\alpha_s(M_Z)= 0.1184(7) $\hfill\cite{Nakamura:2010zzi}								&  $\hat B_{B_d} =
1.26(11)$\hfill\cite{Laiho:2009eu}\\\cline{1-1}
$m_u(2\gev)=(2.1\pm0.1)\mev $ 	\hfill\cite{Laiho:2009eu}						&  $\hat B_{B_s} =
1.33(6)$\hfill\cite{Laiho:2009eu}\\
$m_d(2\gev)=(4.73\pm0.12)\mev$	\hfill\cite{Laiho:2009eu}							& $\hat B_{B_s}/\hat B_{B_d}
= 1.05(7)$ \hfill \cite{Laiho:2009eu} \\
$m_s(2\gev)=(93.4\pm1.1) \mev$	\hfill\cite{Laiho:2009eu}				&
$F_{B_d} \sqrt{\hat
B_{B_d}} = 226(15)\mev$\hfill\cite{Laiho:2009eu} \\
$m_c(m_c) = (1.279\pm 0.013) \gev$ \hfill\cite{Chetyrkin:2009fv}					&
$F_{B_s} \sqrt{\hat B_{B_s}} =
279(15)\mev$\hfill\cite{Laiho:2009eu} \\
$m_b(m_b)=4.19^{+0.18}_{-0.06}\gev$\hfill\cite{Nakamura:2010zzi} 			& $\xi =
1.237(32)$\hfill\cite{Laiho:2009eu}
\\
$m_t(m_t) = 163(1)\gev$\hfill\cite{Laiho:2009eu,Allison:2008xk} &  $\eta_B=0.55(1)$\hfill\cite{Buras:1990fn,Urban:1997gw}  \\
$M_t=172.9\pm0.6\pm0.9 \gev$\hfill\cite{Nakamura:2010zzi} 						&  $\Delta M_d = 0.507(4)
\,\text{ps}^{-1}$\hfill\cite{Nakamura:2010zzi}\\\cline{1-1}
$m_K= 497.614(24)\mev$	\hfill\cite{Nakamura:2010zzi}								&  $\Delta M_s = 17.73(5)
\,\text{ps}^{-1}$\hfill\cite{Abulencia:2006ze,Aaij:2011qx}\\	
$F_K = 156.1(11)\mev$\hfill\cite{Laiho:2009eu}												&
$S_{\psi K_S}= 0.679(20)$\hfill\cite{Nakamura:2010zzi}\\
$\hat B_K= 0.764(10)$\hfill\cite{Laiho:2009eu}												&
$S_{\psi\phi}= 0.0002\pm 0.087$\hfill\cite{Clarke:1429149}\\\cline{2-2}
$\kappa_\epsilon=0.94(2)$\hfill\cite{Buras:2008nn,Buras:2010pza}										&
$\mathcal{B}(B^+\to\tau^+\nu)=(1.64\pm0.34)\times10^{-4}$\hfill\cite{Nakamura:2010zzi}\\	
$\eta_1=1.87(76)$\hfill\cite{Brod:2011ty}												
	& $\tau_{B^\pm}=(1641\pm8)\times10^{-3}\,\text{ps}$\hfill\cite{Nakamura:2010zzi} \\\cline{2-2}		
$\eta_2=0.5765(65)$\hfill\cite{Buras:1990fn}												
&$|V_{us}|=0.2252(9)$\hfill\cite{Nakamura:2010zzi}\\
$\eta_3= 0.496(47)$\hfill\cite{Brod:2010mj}												
& $|V_{cb}|=(40.6\pm1.3)\times
10^{-3}$\hfill\cite{Nakamura:2010zzi}\\
$\Delta M_K= 0.5292(9)\times 10^{-2} \,\text{ps}^{-1}$\hfill\cite{Nakamura:2010zzi}	&
$|V^\text{incl.}_{ub}|=(4.27\pm0.38)\times10^{-3}$\hfill\cite{Nakamura:2010zzi}\\
$|\eps_K|= 2.228(11)\times 10^{-3}$\hfill\cite{Nakamura:2010zzi}					&
$|V^\text{excl.}_{ub}|=(3.12\pm0.26)\times10^{-3}$\hfill\cite{Laiho:2009eu}	\\
\hline
\end{tabular}  }
\caption {\textit{Values of the experimental and theoretical
    quantities used as input parameters.}}
\label{tab:input}~\\[-2mm]\hrule
\end{table}

Having fixed the three parameters of the CKM matrix to the values in (\ref{fixed}), for a given $\vub$  the {``true''} values
of the angle $\beta$  and
of the element $\vtd$
are  obtained from the unitarity of the CKM matrix:
\begin{equation} \label{eq:Rt_beta}
\vtd=\vus \vcb R_t,\quad
R_t=\sqrt{1+R_b^2-2 R_b\cos\gamma} ~,\quad
\cot\beta=\frac{1-R_b\cos\gamma}{R_b\sin\gamma}~,
\end{equation}
where
\be\label{Rb}
 R_b=\left(1-\frac{\lambda^2}{2}\right)\frac{1}{\lambda}\frac{|V_{ub}|}{\vcb}.
\ee

\begin{table}[!tb]
\centering
\begin{tabular}{|c||c|c|c|}
\hline
 & Scenario 1: & Scenario 2:   & Experiment\\
\hline
\hline
  \parbox[0pt][1.6em][c]{0cm}{} $|\varepsilon_K|$ & $1.72(22)  \cdot 10^{-3}$  & $2.28(28)\cdot 10^{-3}$ &$ 2.228(11)\times 10^{-3}$ \\
 \parbox[0pt][1.6em][c]{0cm}{}$(\sin2\beta)_\text{true}$ & 0.623(25) &0.812(23)  & $0.679(20)$\\
 \parbox[0pt][1.6em][c]{0cm}{}$\Delta M_s\, [\text{ps}^{-1}]$ &19.0(21)&  19.1(21) &$17.73(5)$ \\
 \parbox[0pt][1.6em][c]{0cm}{} $\Delta M_d\, [\text{ps}^{-1}]$ &0.55(6) &0.56(6)   &  $0.507(4)$\\
\parbox[0pt][1.6em][c]{0cm}{}$\mathcal{B}(B^+\to \tau^+\nu_\tau)$&  $0.62(14) \cdot 10^{-4}$&$1.19(20)\cdot 10^{-4}$ & $0.99(25) \times
10^{-4}$\\
\hline
\end{tabular}
\caption{\it SM prediction for various observables for  $|V_{ub}|=3.1\cdot 10^{-3}$ and $|V_{ub}|=4.3\cdot 10^{-3}$ and $\gamma =
68^\circ$ compared to experiment. 
}\label{tab:SMpred}~\\[-2mm]\hrule
\end{table}

In Table~\ref{tab:SMpred} we 
summarize for completeness the SM results for $|\varepsilon_K|$,  $\Delta M_{s,d}$, 
$\left(\sin 2\beta\right)_\text{true}$ and $\mathcal{B}(B^+\to \tau^+\nu_\tau)$, obtained from (\ref{eq:Rt_beta}), 
setting
$\gamma = 68^\circ$ and  choosing two values for $\vub$ corresponding to two 
scenarios defined in Section~\ref{sec:1}.
We observe that for both choices of $\vub$ the data show significant deviations from the SM predictions but 
the character of the NP which could cure these tensions depends on the 
choice of $\vub$ as already discussed in detail in \cite{Buras:2012ts} and 
summarized at the beginning of this paper.

What is 
striking in this table is that 
the predicted central values of $\Delta M_s$  and $\Delta M_d$, although 
slightly above the data,  are both in  good agreement with the latter 
when hadronic uncertainties are taken into account. In particular 
the central value of the ratio $\Delta M_s/\Delta M_d$ is 
 very close to  the data:
\be\label{Ratio}
\left(\frac{\Delta M_s}{\Delta M_d}\right)_{\rm SM}= 34.5\pm 3.0\qquad {\rm exp:~~ 35.0\pm 0.3}
\ee
These results depend on the lattice input and in the case 
of $\Delta M_d$ on the value of $\gamma$. Therefore to get a better insight 
both lattice input and the tree level determination of $\gamma$ 
have to improve.

  We also note that in the case of $B^+\to\tau^+\nu_\tau$ the disagreement
of the data with the SM softened significantly with the new
result from Belle Collaboration \cite{BelleICHEP}. The
new world average provided by the UTfit collaboration of
$\mathcal{B}(B^+ \to \tau^+ \nu)_{\rm exp} = (0.99 \pm 0.25) \times 10^{-4}$
\cite{Tarantino:2012mq} given in Table~\ref{tab:SMpred}
is in perfect agreement with the SM in scenario S2 and only by $1.5\sigma$
above the SM value in scenario S1.

\boldmath
\subsection{Results in $MU(2)^3$ Models}
\unboldmath
Let us begin the discussion with the mass differences $\Delta M_{s,d}$. 
The  $MU(2)^3$ models just passed a very important test. The new parameter 
free prediction (\ref{CMFV3}) of these models agrees with data. To bring 
the separate values of $\Delta M_d$ and $\Delta M_s$ closer to the data 
we can just take
\be\label{rB}
r_B=0.93\pm 0.10.
\ee
This is clearly possible in these models because NP contributions 
generating the non-vanishing phase $\varphi_{\rm new}$ can interfere 
destructively with SM contribution. We will see an explicit example in 
the next Section. Moreover, in contrast to CMFV models this NP effect 
has no impact on $\varepsilon_K$. But it should be noted that while we 
got this result for free, in a concrete  dynamical $MU(2)^3$ model  
(\ref{rB}) will constitute a constraint on the fundamental parameters of this 
model.

We have still two parameters to our disposal, $r_K$ and  $\varphi_{\rm new}$. 
$r_K$ enters $\varepsilon_K$, while  $\varphi_{\rm new}$ the CP asymmetries 
 $S_{\psi K_S}$ and $S_{\psi\phi}$. At first sight one could consider these two
parameters independently, but one should notice that $\varepsilon_K$ and 
$S_{\psi K_S}$ depend on the same phase $\beta$, which in turn depends 
sensitively on $\vub$ and very mildly on $\gamma$. Therefore, in order to get the full picture whether 
a given model works or not we have to consider $S_{\psi K_S}$, $S_{\psi\phi}$, 
$\varepsilon_K$, $\vub$ and the two new parameters simultaneously as they 
are correlated with each other when the experimental data are taken into 
account.

\begin{figure}[!tb]
 \centering
\includegraphics[width = 0.6\textwidth]{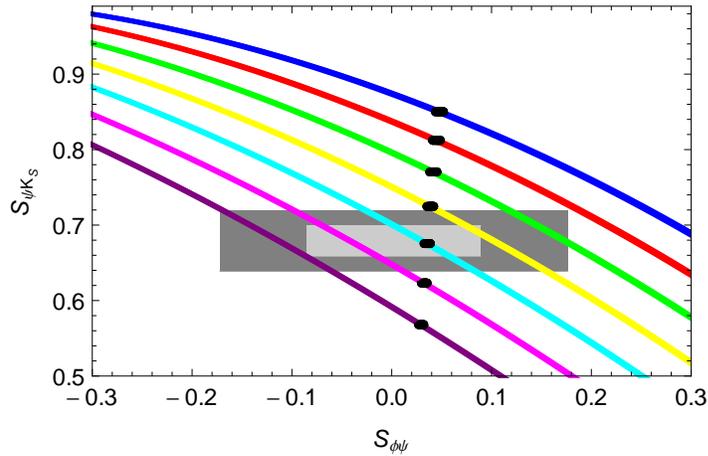}
\caption{ \it $S_{\psi K_S}$ vs. $S_{\psi \phi}$ in  models with 
$U(2)^3$ symmetry for different values of $\vub$ and $\gamma \in[58^\circ, 78^\circ]$. From top to bottom: $\vub =$ $0.0046$ (blue),
$0.0043$ (red), $0.0040$ (green),
$0.0037$ (yellow), $0.0034$ (cyan), $0.0031$ (magenta), $0.0028$ (purple). Light/dark gray: experimental $1\sigma/2\sigma$ region.
}\label{fig:SvsS}~\\[-2mm]\hrule
\end{figure}

In Fig.~\ref{fig:SvsS} we show $S_{\psi K_S}$ vs. $S_{\psi\phi}$ for different 
values of $\vub$ for  $\gamma=68^\circ\pm 10^\circ$.  The light gray (dark gray) area shows 
1$\sigma$ (2$\sigma$) experimental ranges for both asymmetries. The black 
dots represent the case of vanishing new phases $\varphi_{B_q}$, in which case 
the formulae in (\ref{CMFV1}) apply. As expected for this case we observe a strong variation of $S_{\psi K_S}$ with $\vub$ but only tiny $\vub$-dependence in 
$S_{\psi\phi}$. The SM and CMFV are represented here roughly by a black dot on the cyan
$\vub=3.4\times 10^{-3}$ line. 

 We would like to emphasize that the uncertainties in the plot in 
    Fig.~\ref{fig:SvsS} for fixed $\vub$ are very small. 
  Indeed this plot is based on the 
   unitarity of the CKM matrix and the $U(2)^3$ relation between new phases 
   in (\ref{MAIN1}). The very small uncertainty to the value of $\gamma$
   can be understood as follows. $S_{\psi\phi}$
    depends very weakly on $\gamma$ as it is an order $\lambda^2$ effect. 
    $S_{\psi K_S}$ depends also weakly on $\gamma$ because as seen in the unitarity 
    triangle, the angle $\beta$ relevant for $S_{\psi K_S}$  is an orthogonal 
    variable to $\gamma$. The remaining theoretical uncertainties due to 
    QCD-penguins are small
\cite{Fleischer:1999nz,Ciuchini:2005mg,Ciuchini:2011kd,Faller:2008zc,Faller:2008gt,Jung:2009pb,DeBruyn:2010hh,Lenz:2011zz,Boos:2004xp,
Li:2006vq,Gronau:2008cc,Jung} and when both asymmetries will be 
measured precisely, also these uncertainties are expected to be fully 
under control.

\begin{figure}[!tb]
 \centering
\includegraphics[width = 0.6\textwidth]{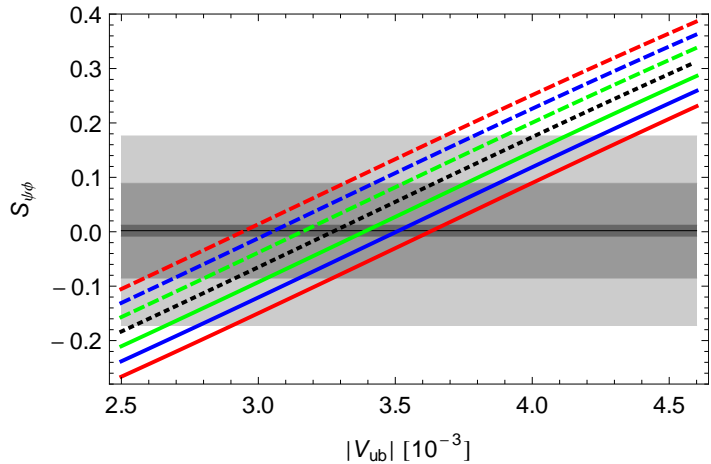}
\caption{\it $S_{\psi \phi}$ vs. $\vub$ in  models with 
$U(2)^3$ symmetry for different values of $S_{\psi K_S}$ (experimental central value and $1$, $2$ and $3\sigma$ values). From top to
bottom: $S_{\psi K_S} = 0.619$ (red dashed), $0.639$ (blue dashed), $0.659$ (green dashed), $0.679$ (black dotted), $0.699$ (green
solid), $0.719$ (blue solid), $0.739$ (red solid). Horizontal black line: experimental central value $S_{\psi\phi}^\text{exp} =
0.002$, dark gray region: $0.002 \pm 0.01$, middle gray: experimental $1\sigma$ range, light gray: experimental
$2\sigma$ range.
}\label{fig:Svsub}~\\[-2mm]\hrule
\end{figure}

In Fig.~\ref{fig:Svsub} we show $S_{\psi\phi}$ vs. $\vub$ in $MU(2)^3$  models for different values of $S_{\psi K_S}$. The information in 
this figure is equivalent to the previous one but it could turn out to be
more useful than the latter once $S_{\psi K_S}$ will be measured very precisely 
at the LHC.

Figs.~\ref{fig:SvsS} and \ref{fig:Svsub} exhibit a number of interesting 
features that we would like to emphasize now.
\begin{itemize}
\item
For values in the ballpark of inclusive determinations and close to the 
central value in (\ref{average}) we find, in agreement with \cite{Barbieri:2011ci}, that 
$S_{\psi\phi}$ is positive and larger than the SM value. The same applies 
to a specific $U(2)^3$ model,  ${\rm 2HDM_{\overline{MFV}}}$ with the dominance 
of flavour blind phases in the Higgs potential, as found 
in \cite{Buras:2010mh,Buras:2010zm,Buras:2012ts,Buras:2012xxx}
\footnote{The operators responsible for NP contributions in this case are 
the $Q^{\rm SLL}_{1,2}$ with the $(S-P)\times(S-P)$ structure but they also 
imply $\varphi_{B_d}=\varphi_{B_s}$. On the other hand in 
 ${\rm 2HDM_{\overline{MFV}}}$ $r_K=1$ to a good approximation.}.
\item
Already now within  $MU(2)^3$ models we find a 1$\sigma$ (2$\sigma$) 
bound 
\be
\vub\le 3.7\times 10^{-3} (4.4\times 10^{-3}).
\ee
\item
 An important new result following from Fig.~\ref{fig:SvsS}
is that, in contrast to  ${\rm 2HDM_{\overline{MFV}}}$, in the $MU(2)^3$ models 
 {\it negative} values of
$S_{\psi\phi}$ can be in principle accommodated so that 
 a future measurement of a negative $S_{\psi\phi}$ would not rule out 
this class of models. However, such a measurement would 
favour in this framework $\vub$ in the ballpark of exclusive determinations.
\item
As seen in particular in Fig.~\ref{fig:Svsub} precise measurements  of  $S_{\psi K_S}$ vs. $S_{\psi \phi}$ would allow in this framework a rather precise determination of $\vub$
subject to much smaller hadronic uncertainties than the usual 
determinations by means of semi-leptonic $B$ decays. Already the 
measurements of 
both asymmetries with errors of $\pm0.01$ would allow the determination 
of $\vub$ with an impressive 
accuracy of $2\%$. However, at this order 
of accuracy the corrections from QCD-penguins to the formulae in (\ref{U21}) 
should be included.
\item
On the other hand once  $S_{\psi K_S}$, $S_{\psi \phi}$ and $\vub$ will 
be determined one day independently of any  $MU(2)^3$ assumptions 
the plots in Figs.~\ref{fig:SvsS} and \ref{fig:Svsub} will allow to test 
the $MU(2)^3$ models as a general framework. If this test will be successfully passed, the selection of the concrete successful $MU(2)^3$
model will require the inclusion of other observables.
\end{itemize}

In the context of the last point, the $\varepsilon_K$ constraint should be 
emphasized. In particular, in the case of a low value of $\vub$, as seen in 
Table~\ref{tab:SMpred}
some enhancement of the function $S_K$ over its SM value 
is required. That is $r_K>1$ which is very natural in
$MU(2)^3$ models.

\begin{figure}[!tb]
 \centering
\includegraphics[width = 0.6\textwidth]{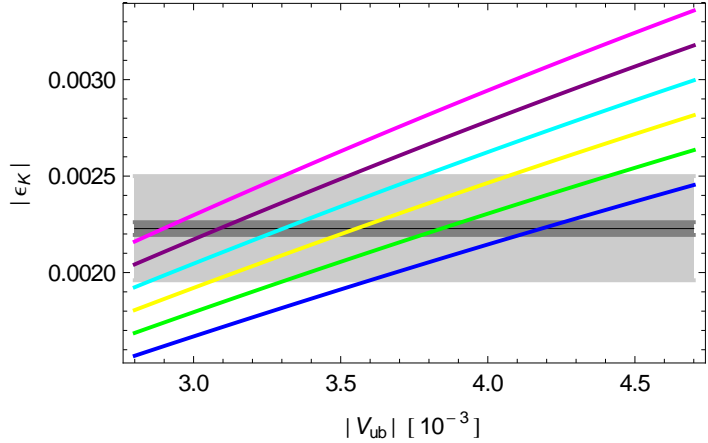}
\caption{\it $|\varepsilon_K|$ vs. $\vub $ in  models with 
$U(2)^3$ symmetry for fixed $S_{\psi K_S}=0.679$ and different 
values of the enhancement factor $r_K$. From top to bottom: $r_K = 1.5$ (magenta), $1.4$ (purple), $1.3$ (cyan), $1.2$ (yellow), $1.1$
(green), $1$ (blue, SM prediction)). Dark gray region: experimental $3\sigma$ range of $|\varepsilon_K|$. Light gray region: theoretical uncertainty in  $|\varepsilon_K|$.
}\label{fig:epsK1}~\\[-2mm]\hrule
\end{figure}

\begin{figure}[!tb]
 \centering
\includegraphics[width = 0.6\textwidth]{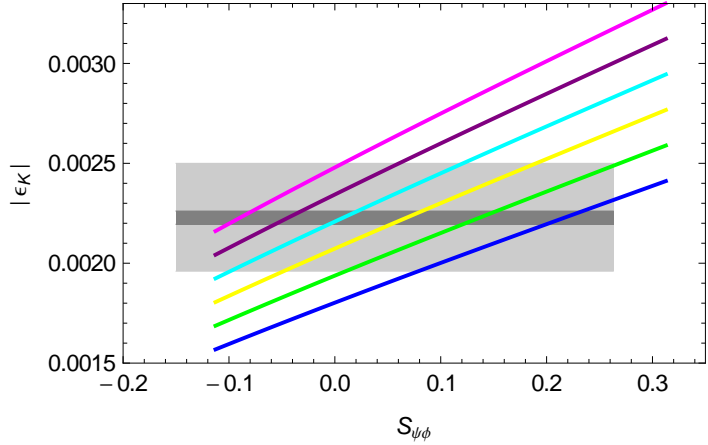}
\caption{\it $|\varepsilon_K|$ vs. $S_{\psi \phi}$ in  models with 
$U(2)^3$ symmetry for fixed $S_{\psi K_S}=0.679$, $\vub\in[0.0028,0.0046]$ and different 
values of the enhancement factor $r_K$ (colours for $r_K$ as in Fig.~\ref{fig:epsK1}). 
}\label{fig:epsSpsiphi}~\\[-2mm]\hrule
\end{figure}

In order to illustrate this point quantitatively, we choose the nominal values of the 
CKM parameters in (\ref{fixed}), set the values of other input parameters 
at their central values in Table~\ref{tab:input} and show in Fig.~\ref{fig:epsK1}
$|\varepsilon_K|$ as a function of $\vub$ for different values of $r_K$. To 
this end we set $S_{\psi K_S}$ at its central experimental value.
As this specifies  uniquely the correlation between $S_{\psi \phi}$ and 
$\vub$, we also show in Fig.~\ref{fig:epsSpsiphi} $|\varepsilon_K|$ as a function of $S_{\psi\phi}$.
The latter correlation is interesting in itself as it shows that in a given 
specific  $MU(2)^3$ model also correlations between $B$-physics and $K$-physics 
are possible.

At this point we should emphasize the difference in the quality of the 
correlations in  Figs.~\ref{fig:SvsS} and \ref{fig:Svsub} on one hand and 
in  Figs.~\ref{fig:epsK1} and \ref{fig:epsSpsiphi} on the other hand. 
While for a fixed value of $\vub$ the theoretical uncertainties in 
 Figs.~\ref{fig:SvsS} and \ref{fig:Svsub} are very small, this is certainly 
not the case of  Figs.~\ref{fig:epsK1} and \ref{fig:epsSpsiphi} in which 
 $|\varepsilon_K|$ is involved. Indeed, as seen in Table~\ref{tab:SMpred} the 
error in the SM prediction for  $|\varepsilon_K|$ amounts to roughly 
$\pm 12\%$. As this error originates dominantly from $\vcb$, $\eta_1$ and 
$\hat B_K$, it is practically independent of $\vub$.  Therefore, in reality 
the values of $r_K$ extracted for fixed values of $\vub$ from these plots have 
an error of roughly $\pm 12\%$, which makes the tests of these correlations 
difficult at present. We show this uncertainty  in  Figs.~\ref{fig:epsK1} and \ref{fig:epsSpsiphi} by incorporating it in the 
experimental value of  $|\varepsilon_K|$.

 Bearing this problem in mind  in the case of a negative $S_{\psi \phi}$ chosen by nature and 
consequently small $\vub$ implied by  $U(2)^3$ symmetry, an enhancement 
of  $|\varepsilon_K|$ by roughly  $20\%$ is required. 

The latter condition is not satisfied in the 
${\rm 2HDM_{\overline{MFV}}}$ model which in the limit of the dominance 
of flavour blind phases in the Higgs potential exhibits $U(2)^3$ symmetry. As in this 
model NP contributions to $\varepsilon_K$ are negligible, this model 
 favours 
$\vub\ge 3.6\cdot 10^{-3}$ \cite{Buras:2012ts,Buras:2012xxx}
and this implies
$S_{\psi\phi}\ge0.05$. 
Finding in the future that nature chooses a {\it negative} value of 
$S_{\psi\phi}$  and/or small (exclusive) value of $\vub$ would put
${\rm 2HDM_{\overline{MFV}}}$ into difficulties. Clear cut conclusion can 
only be reached when theoretical error on  $\varepsilon_K$ will be 
decreased.

Also a decrease of the experimental 
error on $S_{\psi\phi}$ without the change of its central value would be 
problematic for this model.

On the other hand as we have seen  $MU(2)^3$ models can be 
in principle consistent with a negative value of $S_{\psi\phi}$ provided 
\be 
\vub\le 3.3\times 10^{-3}, \qquad S_K\ge 1.20~S_0(x_t)\,,
\ee
with precise values depending on $S_{\psi K_S}$ as seen in Fig.~\ref{fig:Svsub} 
 and uncertainties in $\varepsilon_K$ discussed above.

\begin{figure}[!tb]
 \centering
\includegraphics[width = 0.6\textwidth]{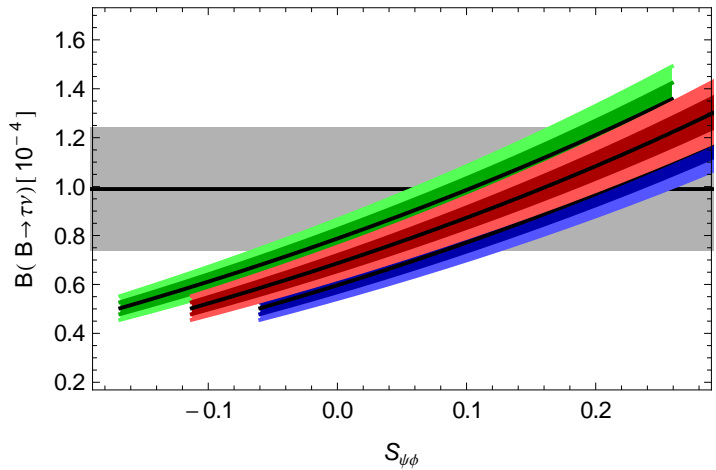}
\caption{\it $\mathcal{B}(B^+\to\tau^+\nu_\tau)$ vs. $S_{\psi \phi}$ in  models with
$U(2)^3$ symmetry for $S_{\psi K_S}=0.679$ (experimental central value, red), $0.719$ (upper $2\sigma$ value, green) and $0.639$
(lower $2\sigma$ value, blue) and $\vub\in[0.0028,0.0046]$
and different values of $F_{B^+} = 190.6~$MeV (central value, black), $(190.6\pm4.6)~$MeV ($1\sigma$ region, dark coloured
region), $(190.6\pm9.2)~$MeV ($2\sigma$ region, light coloured region). We show 
$1\sigma$ experimental range for 
$\mathcal{B}(B^+\to\tau^+\nu_\tau)$ (gray).
}\label{fig:Btaunu}~\\[-2mm]\hrule
\end{figure}

However, as is well known, the branching ratio for $B^+\to\tau^+\nu_\tau$ 
depends sensitively on $\vub$ and this may not allow negative values 
of $S_{\psi\phi}$. As in $U(2)^3$ models there is no obvious possibility to 
enhance the branching ratio in question, we use the SM expression for it 
and show in Fig.~\ref{fig:Btaunu} for fixed $S_{\psi K_S}$ the correlation between
$\mathcal{B}(B^+\to\tau^+\nu_\tau)$ and  $S_{\psi\phi}$ for different values 
of $F_{B^+}$.   We show there also the present world average for 
$\mathcal{B}(B^+\to\tau^+\nu_\tau)$. Evidently we have to wait for improved 
data on the quantities involved but
a {\it negative}  $S_{\psi\phi}$ in this framework has a clear tendency to
imply values of  $\mathcal{B}(B^+\to\tau^+\nu_\tau)$ below the present data.

In this context we would also like to remark that in CMFV the increase of 
$F_{B^+}$ while improving the agreement for  $\mathcal{B}(B^+\to\tau^+\nu_\tau)$ 
would worsen  the agreement of the theory and data for $\Delta M_{s,d}$. 
In the $MU(2)^3$ models this change could  in principle be compensated 
by non-MFV effects in $\Delta M_{s,d}$.

\boldmath
\section{Strategy for Testing $MU(2)^3$ Scenario}\label{sec:5}
\unboldmath
\subsection{Preliminaries}
In order to test any NP scenario through flavour physics we need a
good precision on the CKM parameters from tree-level decays and on the 
non-perturbative parameters entering $\Delta F=2$ observables. Concerning 
the latter, the absence of contributions from new operators does not increase 
the hadronic uncertainties in the  $MU(2)^3$ models in question relatively to SM and CMFV 
models. Moreover, lattice calculations made significant progress in the last 
years \cite{Laiho:2009eu,Bouchard:2011xj,Boyle:2012qb,Bertone:2012cu} and further progress is expected in the coming years. 

The optimal choice for the input parameters beyond those listed in 
(\ref{fixed}) would certainly be $\vub$ determined in tree-level decays 
without any or only small pollution from NP. However, as we have seen the 
present status of $\vub$  is very unsatisfactory and it is likely that 
only after Super-Belle and later Super-B will provide the data and improved theory we will 
know a significantly more precise value of $\vub$. 

However, as pointed out in \cite{Buras:2002yj}, a very efficient method to 
determine the CKM parameters is to use in addition to the set (\ref{fixed}), 
the asymmetry $S_{\psi K_S}$. That is the use of the set $(\gamma,\beta)$ allows 
with a smaller precision on both variables to determine the apex of the Unitarity Triangle $(\bar\rho,\bar\eta)$ than it is required if for instance 
$(\beta,R_t)$ or in particular $(\gamma,R_t)$ are used.  Now $S_{\psi K_S}$ 
is polluted by NP but in  the $MU(2)^3$ scenario this pollution has a very simple 
structure represented by a universal new phase $\varphi_{\rm new}$. Moreover this phase can be 
determined by invoking precisely  $S_{\psi \phi}$. 

Thus we are led to a new strategy for the determination of CKM parameters
 which generally would not be evidently 
efficient but in this simple NP scenario could indeed turn out to be useful 
in the second phase of the LHC. Of course, we are aware of possible QCD penguin 
effects absent in our analysis, that one has to consider in order to reach 
sufficient precision on the extraction of the parameters involved from 
the data on  $S_{\psi K_S}$ and $S_{\psi\phi}$.
However, this issue  is a science for itself
\cite{Fleischer:1999nz,Ciuchini:2005mg,Ciuchini:2011kd,Faller:2008zc,Faller:2008gt,Jung:2009pb,DeBruyn:2010hh,Lenz:2011zz,
Boos:2004xp,Li:2006vq, Gronau:2008cc,Jung} and clearly beyond the scope of our
paper. Our strategy then involves five steps.

\subsection{Strategy in Five Steps}

{\bf Step 1.}

 Use the following experimental input from tree diagram dominated decays
\be
\vus, \quad \vcb,\quad \gamma, \quad S_{\psi K_S}, \quad S_{\psi\phi}.
 \ee
$\vus$ is already precisely known and the accuracy on  $\gamma$ should 
be improved in the coming years by the LHCb. $\vcb$ is known presently 
with an accuracy of $2-3\%$ and improvements are expected from SuperKEKB and 
SuperB.

{\bf Step 2.}

 Using the formulae of Section~\ref{sec:3} together with (\ref{eq:Rt_beta}) 
and (\ref{Rb})
find 
\be
\vub, \quad \varphi_{\rm new}, \quad \beta=\beta_{\rm true}.
\ee

{\bf Step 3.}

Using future improved values of non-perturbative parameters from 
lattice calculations extract
\be
r_K,\qquad r_B.
\ee 
If $r_K$ turns out to be below unity, this NP scenario will be put 
into difficulties, but presently the data favour $r_K$ above one.

 {\bf Step 4.}

Extract $\vub$ from future data and improved theory and compare the 
result with the value implied by Fig.~\ref{fig:SvsS}. 
This is a crucial test of this NP scenario. Assuming that this test is 
passed go to a specific model in the last step.

{\bf Step 5.}

Verify whether a given $MU(2)^3$ model is consistent with the extracted 
values of
\be
r_K, \quad  r_B, \quad \varphi_{\rm new}.
\ee

We will now illustrate this strategy with a specific example.

\subsection{Explicit Example}
The specific supersymmetic model studied in 
\cite{Barbieri:2011ci,Barbieri:2011fc} has precisely the structure of $MU(2)^3$ 
and with respect to $\Delta F=2$ transitions can be
summarized simply as follows:
\be
r_K=1+|\xi_L|^4F_0\,,
\ee
\be\label{equ:rBphase}
r_Be^{2i\varphi_{\rm new}}=1+\xi^2_L F_0 \,,
\ee
where 
\be
\xi_L=\frac{s_Lc_d}{\vts}e^{i\gamma_L}\,,
\ee
with $\gamma_L$ being an arbitrary phase and  $s_L$ and $c_d$  defined in \cite{Barbieri:2011ci,Barbieri:2011fc}. Next
\begin{align}
 F_0 & = \frac{2}{3}\left(\frac{g_s}{g}\right)^4\frac{M_W^3}{m_{Q_3}^2}\frac{1}{S_0(x_t)}\left[f_0(x_g) +
\mathcal{O}\left(\frac{m_{Q_l}^2}{m_{Q_h}^2}\right)\right]\,,\\
f_0(x) & = \frac{11 + 8x-19 x^2+26x\log x + 4 x^2\log x}{3(1-x)^3}\,,\\
x_g & = \frac{m_{\tilde g}^2}{m_{Q_3}^2}\,,\quad f_0(1) = 1\,.
\end{align}
Here $m_{\tilde{g}}$ is the gluino mass and $m_{Q_3}^2$ the third generation squark mass. 
Indeed, this model has three new free parameters
\be
|\xi_L|, \quad F_0, \quad \gamma_L\,.
\ee
In \cite{Barbieri:2011ci} in Fig.~2 it is shown
that for $400~\text{GeV}\lesssim m_{\tilde{g}}\lesssim 1000~$GeV and $500~\text{GeV}\lesssim  m_{Q_3}\lesssim
1200~$GeV one gets $F_0$ roughly between 0.1 and 0.02 and $|\xi_L|$ is of $\mathcal{O}(1)$.

\begin{figure}[!tb]
 \centering
\includegraphics[width = 0.45\textwidth]{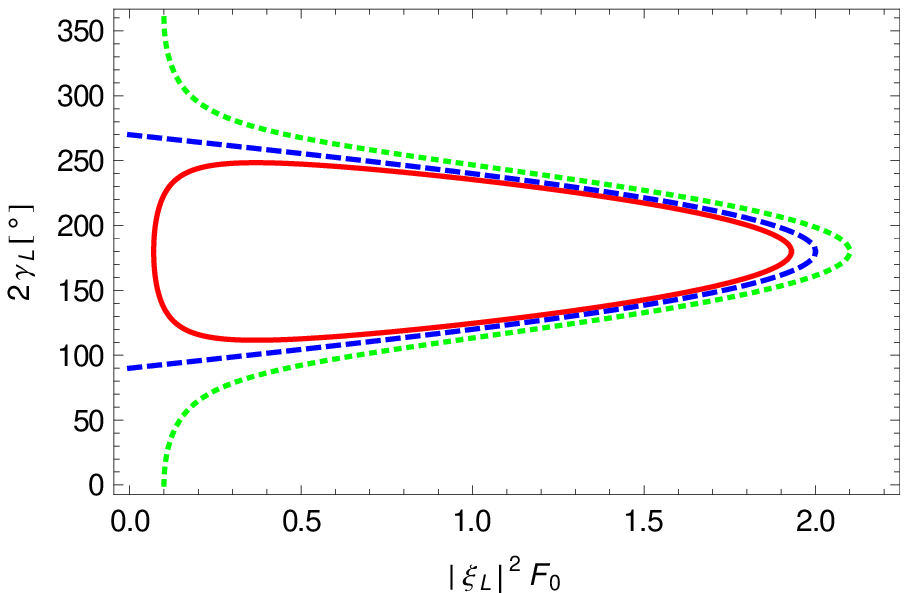}
\includegraphics[width = 0.45\textwidth]{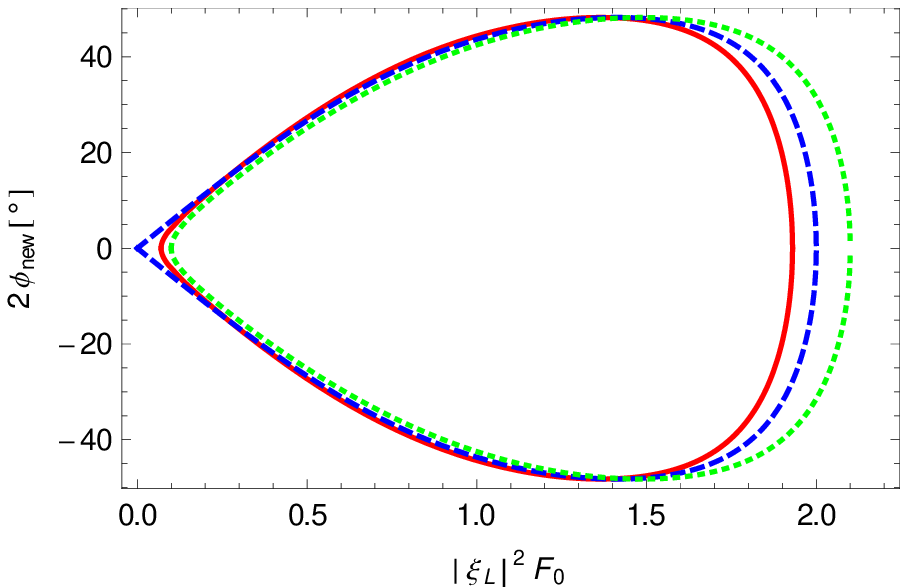}
\includegraphics[width = 0.45\textwidth]{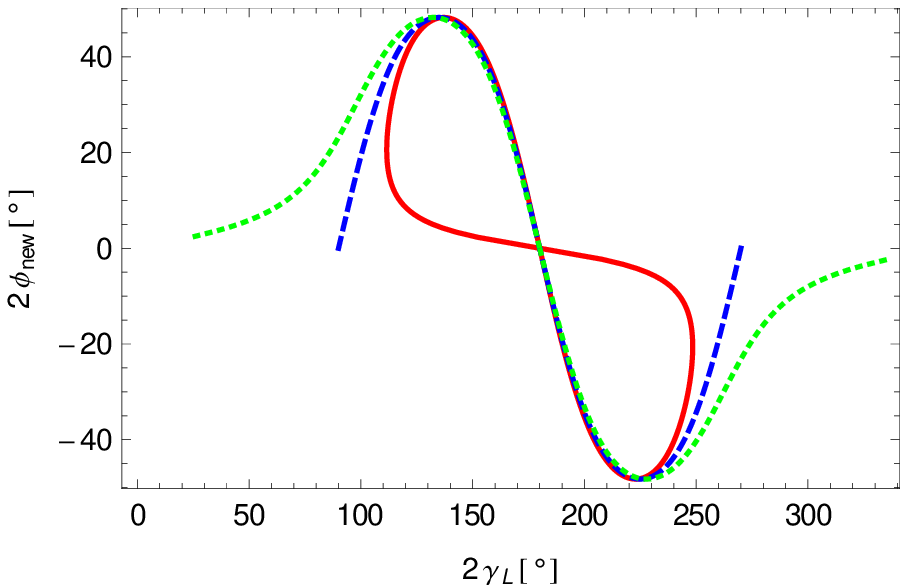}
\caption{\it For fixed $r_B = 0.93$ (red solid), $r_B = 1$ (blue dashed) and $r_B = 1.1$ (green dotted) relation between $2\gamma_L$,
$2\varphi_\text{new}$ and $|\xi_L|^2F_0$.}\label{fig:phases}~\\[-2mm]\hrule
\end{figure}

We follow the steps from the previous section. As inputs we fix $\vus =0.2252$, $\vcb =0.0406$ and $\gamma = 68^\circ$ and use three
different values for $S_{\psi K_S}$ corresponding to its experimental $1\sigma$ range (0.659, 0.679, 0.699) and for $S_{\psi \phi}$ we use
$-0.2$, $-0.1$, 0, 0.1 and 0.2. This is listed in the first two columns of Table~\ref{tab:1}.
As already visualized in Fig.~\ref{fig:SvsS} and~\ref{fig:Svsub} with this information we can determine $\vub$  and also
$2\varphi_\text{new}$ and $\beta_\text{true}$. The results are written in columns 3--5 in Table~\ref{tab:1}. As a next step we want to shift
$|\varepsilon_K|$ to its measured value $2.228\cdot 10^{-3}$ allowing for a 12\% error on $r_K$ to account for the uncertainties in
$|\varepsilon_K|$. This leads to the values of $r_K$ quoted in Tab.~\ref{tab:1}. Although there is no direct correlation between $K$ and
$B_{d,s}$ sector in $U(2)^3$ models we get a connection due to the dependence of $|\varepsilon_K|$ on $\beta_\text{true} =
\beta_\text{true}(\vub)$. Fixing $r_B$ in our explicit example we get a one-to-one relation between $|\xi_L|^2F_0$ and $2\gamma_L$ and
between $|\xi_L|^2F_0$ and $2\varphi_\text{new}$ and thus also between $2\gamma_L$ and $2\varphi_\text{new}$. This is shown for $r_B =
0.93$, $r_B = 1$ and $r_B = 1.1$ in Fig.~\ref{fig:phases}. In the following we set $r_B = 0.93$ as suggested by current lattice values and
calculate the corresponding model parameter $|\xi_L|^2F_0$ using Eq.~(\ref{equ:rBphase}). Using this and $r_K$ we can derive a range for
$|\xi_L|^2$ and then also for $F_0$ (see last three columns of Table~\ref{tab:1}). But $|\xi_L|^2$ and $F_0$ have to be combined such that
one gets the right $|\xi_L|^2F_0$ from the 7th column (e.g. the lower bound from $|\xi_L|^2$ and the upper bound from $F_0$). The derived
values of $F_0$ can now be compared with the range 0.02--0.10 that is obtained for reasonable gluino and sbottom masses. For $F_0\leq 0.02$
one needs rather heavy gluino and sbottom masses. For $S_{\psi\phi}=0.2$ where $\vub$ is near the inclusive value one needs very light
gluino and sbottom masses near 300~GeV. The limiting factor for $S_{\psi\phi} = -0.2$ in this model are not the SUSY masses but the small
value of $\vub\approx 0.0024$. However $S_{\psi\phi} = -0.1$ could still be reached for reasonable gluino and sbottom masses in case of
$\vub\approx 0.0028$.

\begin{table}[!htb]
 \center{\begin{tabular}{|c|c||ccc||c|ccc|}
          \hline
$S_{\psi K_S}$ & $S_{\psi\phi}$ & $\vub$& $2\varphi_\text{new}$& $\beta_\text{true}$& $r_K$&$|\xi_L|^2
F_0$&$|\xi_L|^2$&$F_0$\\
\hline\hline
  \parbox[0pt][1.6em][c]{0cm}{}0.679 & 0.2 & 0.0041 &-9.1$^\circ$ &25.9$^\circ$ &$1.02\pm0.12$&0.169&[0,0.83]&$\geq 0.204$\\
  \parbox[0pt][1.6em][c]{0cm}{}0.679 & 0.1 & 0.0037 &-3.5$^\circ$ &23.2$^\circ$ &$1.15\pm0.14$&0.092&[0.11,3.17]&[0.029,0.838]\\
  \parbox[0pt][1.6em][c]{0cm}{}0.679 & 0 & 0.0033& 2.0$^\circ$&20.4$^\circ$&$1.31\pm0.16$&0.078&[1.93,6.05]&[0.013,0.040]\\
  \parbox[0pt][1.6em][c]{0cm}{}0.679 & -0.1 & 0.0028& 7.5$^\circ$&17.7$^\circ$&$1.52\pm0.18$&0.145&[2.35,4.83]&[0.030,0.062]\\
  \parbox[0pt][1.6em][c]{0cm}{}0.679 & -0.2 & 0.0024 &13.0$^\circ$ &14.9$^\circ$ &$1.81\pm0.22$&0.233&[2.53,4.42]&[0.053,0.092]\\
\hline
  \parbox[0pt][1.6em][c]{0cm}{}0.659 & 0.2 &0.0040 & -9.1$^\circ$&25.2$^\circ$&$1.05\pm0.13$&0.169&[0,1.06]&$\geq 0.159$\\
  \parbox[0pt][1.6em][c]{0cm}{}0.659 & 0.1 &0.0036 & -3.6$^\circ$&22.4$^\circ$&$1.19\pm0.14$&0.093&[0.54,3.56]&[0.026,0.172]\\
  \parbox[0pt][1.6em][c]{0cm}{}0.659 & 0 & 0.0032& 1.9$^\circ$&19.7$^\circ$&$1.36\pm0.16$&0.077&[2.60,6.76]&[0.011,0.030]\\
  \parbox[0pt][1.6em][c]{0cm}{}0.659 & -0.1 & 0.0027&7.4$^\circ$ &16.9$^\circ$&$1.59\pm0.19$&0.143 &[2.79,5.44]&[0.026,0.051]\\
  \parbox[0pt][1.6em][c]{0cm}{}0.659 & -0.2 &0.0023 & 12.9$^\circ$&14.1$^\circ$&$1.90\pm0.23$&0.231&[2.90,4.89]&[0.047,0.080]\\
\hline
  \parbox[0pt][1.6em][c]{0cm}{}0.699 & 0.2 & 0.0042& -9.0$^\circ$&26.7$^\circ$&$0.99\pm0.12$&0.168&[0,0.66]&$\geq 0.256$\\
  \parbox[0pt][1.6em][c]{0cm}{}0.699 & 0.1 & 0.0038& -3.5$^\circ$&23.9$^\circ$&$1.11\pm0.13$&0.092&[0,2.62]&$\geq 0.035$\\
  \parbox[0pt][1.6em][c]{0cm}{}0.699 & 0 & 0.0034& 2.0$^\circ$&21.2$^\circ$&$1.26\pm0.15$&0.078&[1.42,5.28]&[0.015,0.055]\\
  \parbox[0pt][1.6em][c]{0cm}{}0.699 & -0.1 & 0.0030&7.5$^\circ$ &18.4$^\circ$&$1.46\pm0.18$&0.145&[1.93,4.42]&[0.033,0.075]\\
  \parbox[0pt][1.6em][c]{0cm}{}0.699 & -0.2 & 0.0025& 13.1$^\circ$&15.6$^\circ$&$1.72\pm0.21$&0.235&[2.17,3.96]&[0.059,0.108]\\

\hline\hline
         \end{tabular}}
\caption{\it Explicit $MU(2)^3$ SUSY example. Column 1--5 correspond to step 1 and 2 where for different inputs of $S_{\psi K_S}$ and
$S_{\psi\phi}$ we determine the corresponding $\vub$, $2\varphi_\text{new}$ and $\beta_\text{true}$. $r_K$ is derived such that
$|\varepsilon_K| = 2.228\cdot 10^{-3}$ with a 12\% error on $r_K$. The corresponding model parameters are listed in column 7--9 using $r_B =
0.93$. For more details see text.}\label{tab:1}~\\[-2mm]\hrule
\end{table}

\section{Rare Decays}\label{sec:6}
\subsection{Preliminaries}
Here we just want to describe briefly the pattern of flavour violation 
in $MU(2)^3$ models in rare $B$ and $K$ decays and compare it to the one in CMFV models. 

First we would like to emphasize  
that the flavour physics in the $MU(2)^3$ models can be 
parametrized, similar to the LHT model, by a set of master functions $F_i$
 ($i=K,d,s$):
\begin{equation}\label{eq310}
S_i\equiv|S_i|e^{i\theta_S^i},
\quad
X_i \equiv \left|X_i\right| e^{i\theta_X^{i}}, \quad
Y_i \equiv \left|Y_i\right| e^{i\theta_Y^i}, \quad
Z_i \equiv \left|Z_i\right| e^{i\theta_Z^i}\,,
\end{equation}

\begin{equation} \label{eq32}
E_i \equiv \left|E_i\right| e^{i\theta_{E}^i}\,,\quad
D'_i \equiv \left|D_i'\right| e^{i\theta_{D'}^i}\,, \quad
E'_i \equiv \left|E_i'\right| e^{i\theta_{E'}^i}\,,
\end{equation}
with the following properties implied by $U(2)^3$ symmetry:
\be
|F_d|=|F_s|, \quad \theta_F^{K}=0, \quad \theta_F^{d}=\theta_F^{s}\equiv\theta_F
\ee
that allow to distinguish these models not only from CMFV models but also 
models 
with non-MFV sources like LHT and Randall-Sundrum models. The information 
which function enters which decay can be found in
\cite{Buras:2003jf}.

 On the other hand as discussed in \cite{Barbieri:2011fc}, contrary 
to $\Delta F=2$ transitions, where the pattern of deviations from the SM 
is unambiguously dictated by $U(2)^3$ symmetry, the predictions for 
$\Delta F=1$ processes are more model dependent. In particular, while 
the quark transitions, similarly to MFV, are governed by the CKM matrix 
and left-handed currents, the lepton currents in FCNC processes 
could in addition to left-handed currents have right-handed component 
absent in the SM and CMFV. Yet, independently of these new operators 
the universality of NP contributions to $B_d$ and 
$B_s$ decays implied by the $U(2)^3$ symmetry is intact and in what follows 
we will only discuss its consequences below.

\subsection{Rare $B$ Decays}
In view of the universality of NP contributions to $B_d$ and 
$B_s$ decays in question,
 several CMFV relations involving only $B_{d,s}$ observables
are not modified. In particular,
\begin{equation}\label{CMFV4}
\frac{\mathcal{B}(B\to X_d\nu\bar\nu)}{\mathcal{B}(B\to X_s\nu\bar\nu)}=
\left|\frac{V_{td}}{V_{ts}}\right|^2
\end{equation}
\begin{equation}\label{CMFV5}
\frac{\mathcal{B}(B_d\to\mu^+\mu^-)}{\mathcal{B}(B_s\to\mu^+\mu^-)}=
\frac{\tau({B_d})}{\tau({B_s})}\frac{m_{B_d}}{m_{B_s}}
\frac{F^2_{B_d}}{F^2_{B_s}}
\left|\frac{V_{td}}{V_{ts}}\right|^2
\end{equation}
remain valid. Moreover, combining (\ref{CMFV3}) and (\ref{CMFV5}) 
one obtains
\cite{Buras:2003td}  
\be\label{CMFV6}
\frac{\mathcal{B}(B_{s}\to\mu^+\mu^-)}{\mathcal{B}(B_{d}\to\mu^+\mu^-)}
=\frac{\hat B_{B_d}}{\hat B_{B_s}}
\frac{\tau( B_{s})}{\tau( B_{d})} 
\frac{\Delta M_{s}}{\Delta M_{d}}, 
\ee
that does not 
involve $F_{B_q}$ and CKM parameters and consequently contains 
smaller hadronic and parametric uncertainties than the formulae considered 
above. It involves
only measurable quantities except for the ratio $\hat B_{s}/\hat B_{d}$
that is already now known  from lattice calculations 
with respectable precision \cite{Shigemitsu:2009jy,Laiho:2009eu} and this 
precision will be improved in the coming years.

The formulae in (\ref{CMFV5}) imply that
\be\label{MAIN3}
\left(\frac{\mathcal{B}(B_s\to\mu^+\mu^-)}{\mathcal{B}(B_d\to\mu^+\mu^-)}\right)_{{\rm MU(2)^3}}=
\left(\frac{\mathcal{B}(B_s\to\mu^+\mu^-)}{\mathcal{B}(B_d\to\mu^+\mu^-)}\right)_{\rm SM}.
\ee

Now, the most recent data from LHCb tell us that the nature does 
not allow for large enhancements of these branching 
ratios.
Indeed the most recent upper bounds from LHCb, ATLAS and CMS at $95\%$ C.L. read \cite{Aaij:2012ac,LHCbBsmumu}
\be\label{LHCb2}
\mathcal{B}(B_{s}\to\mu^+\mu^-) =(3.2^{+1.5}_{-1.2}) \times 10^{-9}, \quad
\mathcal{B}(B_{s}\to\mu^+\mu^-)^{\rm SM}=(3.23\pm0.27)\times 10^{-9},
\ee
\be\label{LHCb3}
\mathcal{B}(B_{d}\to\mu^+\mu^-) \le  9.4\times 10^{-10}, \quad
\mathcal{B}(B_{d}\to\mu^+\mu^-)^{\rm SM}=(1.07\pm0.10)\times 10^{-10}.
\ee
We have shown also SM predictions for these 
observables \cite{Buras:2012ru}.  The  corrections from $\Delta\Gamma_s$, 
pointed out in  \cite{deBruyn:2012wj,deBruyn:2012wk}, are not taken into 
account in these values. We will comment on them below.

The distinction between CMFV and $MU(2)^3$ models on the basis of $\Delta B = 1$ decays can only be made through 
$B$ observables in which the universal phase $\varphi_{\rm new}$ matters. 
These are for instance CP-asymmetries in $B\to X_{s,d}\gamma$, $B\to K^*(\varrho)\gamma$ 
and certain  angular observables in  $B\to K^*\ell^+\ell^-$. Such an 
analysis in a supersymmetric framework with $U(2)^3$ symmetry has 
been presented in  \cite{Barbieri:2011fc}. 

 Here we would like to point out  that the decays
$B_{s,d}\to\mu^+\mu^-$ offer an additional test of the $MU(2)^3$ models 
related to CP violation.
Indeed, 
as pointed out in \cite{deBruyn:2012wj,deBruyn:2012wk}  it 
is possible to define  CP-asymmetries 
$S^{s,d}_{\mu^+\mu^-}$ \cite{deBruyn:2012wk} analogous to $S_{\psi K_S}$ and 
$S_{\psi\phi}$ that measure the phases of the functions $Y_{s,d}$ in 
(\ref{eq310}).

The authors of \cite{deBruyn:2012wk} provide a general expression for 
$S^{s,d}_{\mu^+\mu^-}$ as functions of Wilson coefficients involved. Using 
this formula we find that in the $MU(2)^3$ models these asymmetries
are simply given as follows
\be\label{Fleischer}
S^s_{\mu^+\mu^-}=S^d_{\mu^+\mu^-}=\sin (2\theta_Y-2\varphi_{\rm new}).
\ee
This simplicity reflects
the fact that scalar operators are absent in the models in question. 
In the SM and CMFV models these asymmetries vanish. As
$S^{s,d}_{\mu^+\mu^-}$ are theoretically clean they should offer a good test 
of $MU(2)^3$ models and of any model with new CP-violating phases.
Finally, we should remark that the correction factor to the branching ratio
$\mathcal{B}(B_s\to\mu^+\mu^-)$ of $9\%$ calculated in \cite{deBruyn:2012wk} 
applies not only to the SM but also to all CMFV models. In $MU(2)^3$ models 
it is modified for $\theta_Y\not=0$ and this modification is governed 
by $\cos (2\theta_Y-2\varphi_{\rm new})$.   Thus only in the case of a large 
$S^s_{\mu^+\mu^-}$ it 
will differ visibly from the SM estimate.

When the corrections from $\Delta\Gamma_s$, 
 pointed out in  \cite{deBruyn:2012wj,deBruyn:2012wk}, are taken into account 
and removed from the data 
the experimental result in (\ref{LHCb2}) is reduced by $9\%$ and 
we find 
\be\label{LHCb2corr}
\mathcal{B}(B_{s}\to\mu^+\mu^-)_{\rm corr} =(2.9^{+1.4}_{-1.1}) \times 10^{-9}, 
\ee
that should be compared with the SM result in (\ref{LHCb2}). While the central 
theoretical value agrees very well with experiment, the large experimental 
error still allows for  NP contributions. 

The result in (\ref{LHCb2corr}) combined with the relation (\ref{CMFV6})
allows now to predict  $\mathcal{B}(B_{d}\to\mu^+\mu^-)$ within  
 CMFV and $MU(2)^3$ models. Using the experimental values for $\Delta M_{s,d}$ 
and $\tau( B_{q})$ and the lattice values for $\hat B_{B_q}$ in 
Table~\ref{tab:input} we find
 \be\label{boundMFV1}
 \mathcal{B}(B_{d}\to\mu^+\mu^-)=(1.0^{+0.5}_{-0.3})\times 10^{-10}, \quad
 ({\rm CMFV,~MU(2)^3}),
 \ee
implying that this result can only be tested after the upgrade of the LHCb 
experiment.

\subsection{Rare $K$ Decays}
 Here to a large extent the pattern of flavour violation is the same as in CMFV models. 
In particular, the correlations between the following observables and decays
\be\label{ClassK}
\varepsilon_K, \quad \kpn, \quad \klpn, \quad  \epe
\ee
is the same as in CMFV models. For non-leptonic transitions this follows 
from the left-handed structure of flavour violating quark currents. For 
channels with neutrinos in the final states we expect that only 
left-handed couplings are relevant. On the other hand in
$K_L\to\mu^+\mu^-$ and $K_L\to \pi^0 \ell^+\ell^-$ right-handed currents 
in the leptonic sector could enter but this depends on the lepton part 
of a given model. Finally,
as the $MU(2)^3$ symmetry does not 
imply any correlations between rare $K$ and $B$ decays, NP effects in 
rare $K$ decays are generally not constrained by rare $B_{s,d}$ decays, 
although in specific $MU(2)^3$ models this could be the case. In this context
one should recall  stringent correlations between $K\to \pi\nu\bar\nu$ 
and $B\to X_{s,d}\nu\bar\nu$ as well as $B\to K^*(K)\nu\bar\nu$ decays 
in CMFV models. These correlations are generally absent here.

\section{Summary and Conclusions}\label{sec:7}
In the present paper we have analyzed correlations between flavour observables 
implied by a global $U(2)^3$ symmetry concentrating on $MU(2)^3$ 
models that are analogous to CMFV models. Indeed, while CMFV models are the 
simplest models with MFV, the $MU(2)^3$  models are probably the simplest 
models with non-MFV interactions.
As the simple, but non-trivial, structure 
of these correlations has been already summarized systematically in previous 
sections we emphasize here only the most important findings.
\begin{itemize}
\item
A global $U(2)^3$ symmetry implies a stringent correlation between 
$S_{\psi K_S}$,  $S_{\psi \phi}$ and $\vub$ that is displayed for $\gamma = 68^\circ\pm 10^\circ$ in 
Fig.~\ref{fig:SvsS}. This correlation constitutes an important 
test of this NP scenario but to our knowledge has not been discussed 
so far in the literature.
\item
$\vub$ can be determined  by 
means of future  precise measurements of $S_{\psi K_S}$ and $S_{\psi \phi}$  
that are subject to significantly smaller uncertainties than contained 
in present determinations of $\vub$ by means of semileptonic B decays. 
As  $S_{\psi K_S}$ is presently best known among these three observables we 
have shown in 
Fig.~\ref{fig:Svsub} the correlation between  $S_{\psi \phi}$ and $\vub$. 
This correlation demonstrates clearly that {\it negative} values 
of  $S_{\psi \phi}$ can be accommodated in this framework provided 
$\vub$ is sufficiently low.
\item
 A low value of $\vub$ implies that in the SM, $|\varepsilon_K|$ is significantly 
below the data. Therefore a given $MU(2)^3$ model with negative or very 
small $S_{\psi\phi}$ must have significant enhancement of  $|\varepsilon_K|$ 
relative to the SM in order to agree with data. This we have shown in 
Figs.~\ref{fig:epsK1} and \ref{fig:epsSpsiphi}.  Unfortunately, as 
seen there, the existing theoretical and parametric uncertainties in the SM evaluation 
of  $|\varepsilon_K|$ do not allow a precise determination of the necessary 
enhancement factor $r_K$ at present.
\item
The correlation between 
 $\mathcal{B}(B^+\to\tau^+\nu_\tau)$ and  $S_{\psi\phi}$ 
given in Fig.~\ref{fig:Btaunu}  will
in due time constitute another important constraint 
on $MU(2)^3$ models implying that future experimental {\it negative} values
of  $S_{\psi\phi}$ would in this framework predict
$\mathcal{B}(B^+\to\tau^+\nu_\tau)\le 0.8\times 10^{-4}$.
\item
We have proposed a five step procedure for the determination of 
free parameters in $MU(2)^3$ that should be useful before $\vub$ will 
be better known.
We have illustrated this procedure by means of a specific supersymmetric  
$MU(2)^3$ model \cite{Barbieri:2011ci}.
\item
In the case of CP-conserving observables several CMFV relations between 
observables involving only $B_d$ and $B_s$ decays are the same as in the SM 
and CMFV models. In the case of $K$ decays this is the case for both CP-violating and CP-conserving observables. Therefore an important distinction between 
CMFV models and  $MU(2)^3$ models will also be offered by looking 
at CMFV relations between $K$ and $B_{s,d}$ flavour observables that are generally violated in  $MU(2)^3$ models.
\item
 Finally we  pointed out that 
$B_{s,d}\to\mu^+\mu^-$ allow to 
test new CP-violating phases in $MU(2)^3$ models with the help of 
CP-asymmetries $S^{s,d}_{\mu^+\mu^-}$ proposed in 
\cite{deBruyn:2012wj,deBruyn:2012wk}. The formula for  $S^{s,d}_{\mu^+\mu^-}$  in $MU(2)^3$ models turns out to be very simple and is given in
(\ref{Fleischer}).
\end{itemize}

We are looking forward to improved experimental data  and improved lattice 
calculations. The correlations identified in this paper will allow to 
monitor how  the attractive and simple $MU(2)^3$ 
models face the future precision flavour data.

{\bf Acknowledgements}\\
We would like to thank Gino Isidori for discussions.
This research was done and financed in the context of the ERC Advanced Grant project ``FLAVOUR'' (267104).
AJB would like to thank the organizers of the ``Origin of Mass 2012'' in
 Nordita Stockholm for their hospitality which was very helpful in completing
 the final steps of this work.

\bibliographystyle{JHEP}
\bibliography{allrefs}
\end{document}